\documentclass{emulateapj}

\usepackage{graphicx}
\usepackage{amssymb}
\usepackage{epstopdf}
\usepackage{amsmath}

\begin{document}

\title{PTF\,12gzk - A Rapidly Declining, High-Velocity Type Ic Radio Supernova}

\begin{abstract}

Only a few cases of type Ic supernovae (SNe) with high-velocity ejecta 
have been discovered and studied. Here we present our analysis of radio and X-ray observations of
a Type Ic SN, PTF\,12gzk. The radio emission rapidly declined less than 10 days after
explosion, suggesting SN ejecta expanding at high velocity ($\sim
0.3~c$). The radio data also indicate that the density of the circumstellar
material (CSM) around the supernova is lower by a factor of $\sim
10$ than the CSM around normal Type Ic SNe. Our observations of this
rapidly declining radio SN at a distance of $58$\,Mpc demonstrates the potential to detect 
many additional radio SNe, given the new capabilities of the VLA (improved sensitivity and
dynamic scheduling), that are currently missed, leading to a biased view of radio SNe Ic. Early
optical discovery followed by rapid radio
observations would provide a full description of the ejecta velocity
distribution and CSM densities around stripped massive star explosions, as well as strong clues
about the nature of their progenitor stars.
\end{abstract}

\author{Assaf Horesh\altaffilmark{1},
  Shrinivas R. Kulkarni\altaffilmark{1},
Alessandra Corsi\altaffilmark{2},
Dale, A. Frail\altaffilmark{3},
S. Bradley Cenko\altaffilmark{4}, 
Sagi Ben-Ami\altaffilmark{5}
Avishay Gal-Yam\altaffilmark{5}, 
Ofer Yaron\altaffilmark{5}, 
Iair Arcavi\altaffilmark{5}, 
Mansi M. Kasliwal\altaffilmark{6}, 
Eran O. Ofek\altaffilmark{5}
}

\altaffiltext{1}{Cahill Center for Astrophysics, California Institute
  of Technology, Pasadena, CA, 91125, USA.}
\altaffiltext{2}{Department of Physics, The George Washington University, 725 21st St, NW, Washington, DC 20052}
\altaffiltext{3}{National Radio Astronomy Observatory, P.O. Box 0, Socorro, NM 87801, USA}
\altaffiltext{4}{Department of Astronomy, University of California, Berkeley, CA 94720-3411, USA.}
\altaffiltext{5}{Department of Particle Physics and Astrophysics, The Weizmann Institute of Science, Rehovot 76100, Israel.}
\altaffiltext{6}{Carnegie Institution for Science, Department of
  Terrestrial Magnetism, 5241 Broad Branch Road, Washington, DC 20008
  USA.}
\altaffiltext{7}{Physics Division, Lawrence Berkeley National
  Laboratory, Berkeley, CA 94720, USA.}
\altaffiltext{8}{Department of Physics (Astrophysics), University of Oxford, DWB, Keble Road, Oxford, OX1 3RH, UK.}

\section{Introduction}

There are two diagnostics provided by radio emission from supernovae.
First, radio (and X-ray) emission is generated as fast-moving ejecta
collide with the circumstellar medium (CSM; Chevalier
1982; Chevalier 1998; Weiler et al. 2002; Chevalier \& Fransson 2006). Thus radio and X-ray observations
can be used to measure the density of optically thin CSM. Next,
radio and X-ray emission can be used to trace the {\it fastest} moving ejecta. In 
contrast, the optical emission is reflective of the bulk of the ejecta, where the optical depth
to visible-light photons is determined, which is
necessarily of lower velocity.  From measuring the run of the density one can infer the ratio of the 
mass loss rate to the escape velocity (wind velocity)
of the progenitor star -- the mass loading parameter. Radio observations are in good accord with
the picture that Type Ib/Ic supernovae arise from stripped massive star progenitors.

Both these diagnostics came to the fore for supernovae associated
with Gamma Ray Bursts (GRBs). The first example was SN\,1998bw (a broad-lined Type Ic; Ic-BL), associated
with the nearby, low-luminosity gamma-ray burst GRB\,980425
(Galama et al. 1998).
Radio observations established that the fastest moving ejecta were
mildly relativistic {\it and} carried a significant amount of the explosion
energy (Kulkarni et al. 1998). The same approach showed that
SN\,2009bb (Type Ic-BL) also had mildly relativistic ejecta and it has been reasonably argued that
either SN\,2009bb did not produce gamma-rays or that the associated GRB was likely not pointed towards us (or we missed
detecting the gamma-rays; Soderberg et al. 2010). It is worth noting that in both cases
the optical line features were very broad, i.e., $0.1\,c$. This discussion naturally 
raises the issue whether there exists supernovae which are intermediate (in
the above two senses) between ordinary Ib/Ic supernovae and SN\,1998bw and
SN\,2009bb. 

Here, we report on radio observations of PTF\,12gzk, a Type Ic supernova
which showed high-velocity optical features and a luminous optical emission
(indicative of large radioactive $^{56}$Ni yield), but without detection of gamma-ray
emission. 
PTF\,12gzk turned out to be a fast-evolving radio SN, but with a lower
shockwave velocity (inferred from the radio emission) and energy than SN\,2009bb. At the same time, the
velocities inferred from the optical spectra resemble those of
SN\,1998bw and SN\,2009bb. This raises the possibility that PTF\,12gzk
is an intermediate object, connecting normal SNe Ic with GRB-SNe
(as discussed above).

In $\S 2$ we describe our observation and analyze them
in $\S 3$. We then discuss the nature of
PTF\,12gzk in $\S 4$. We summarize our results and discuss their
implications on future studies in $\S 5$. 


\section{Multi-wavelength Observations}

\subsection{The Optical Discovery of PTF\,12gzk}

The Palomar Transient Factory  (PTF; Law et al. 2009; Rau et al. 2009) discovered a peculiar SN Ic,
PTF\,12gzk (Ben-Ami et al. 2012a), as reported in
in Ben-Ami et al. (2012b). In short, PTF\,12gzk was
discovered on 2012 July 24 UTC in the galaxy SDSS J221241.53+003042.7, $58$ Mpc
away ($z=0.0137$). The SN was not detected in previous images taken by PTF on July
19, down to a $3\sigma$ limiting magnitude of 20.6\,mag. The SN was
detected at a magnitude of $m=20.1$\,mag in $r$ band and reached
a peak magnitude of $m=15$\,mag ($M=-19$) on August 7, 2012.  Based on
the optical rise time Ben-Ami et al. (2012b) estimate the explosion
time to be $t_{0} = $ 2012 Jul
24.0 $\pm 1$\,d. Followup
optical data suggest that PTF\,12gzk is a peculiar Type Ic
supernova, with high expansion velocities, long rise time and a large
radioactive $^{56}$Ni mass, pointing towards an energetic explosion. 
PTF follow-up programs were rapidly triggered (Gal-Yam et al. 2011) including 
the {\it Swift} satellite, the
Very Large Array (VLA), and the Combined Array for Research in
Millimeter-wave Astronomy (CARMA) to observe PTF\,12gzk in X-ray,
cm-wavelength, and mm-wavelength, respectively. Additional follow-up
observations and analysis (including {\it HST} UV spectroscopy) are presented in Ben-ami et al. 
2013 (in preparation).

\subsection{Radio Observations}

We observed PTF\,12gzk using VLA\footnote{VLA
  program 12A-363 (PI Horesh)} on UTC 2012 August 1. The
observation was undertaken in C
(6\,GHz) band using a 2\,GHz bandwidth. The VLA configuration at
that time was the B
configuration. In our observation we used J$2215-0335$ as a phase calibrator and 3C48
as a flux calibrator. The data were reduced using the AIPS\footnote{http://www.aips.nrao.edu} software.

We detect a source with a flux of $77\pm 11$\,$\mu$Jy at the position $\alpha = 22^{\mathrm{h}} 12^{\mathrm{m}} 41.55^{\mathrm{s}}, \delta = 00^{\circ} 30\arcmin 43\farcs1$, consistent with the optical position of the SN within
$0\farcs 5$. The radio luminosity is thus $L_{\nu}=3\times 10^{26}$ erg s$^{-1}$
Hz$^{-1}$. Splitting the data into two sub-band frequencies we
measured a flux of $79\pm 16$\,$\mu$Jy and $82\pm 14$\,$\mu$Jy at $4.8$\,GHz
and $7.4$\,GHz, respectively, implying a spectrum with a power-low index of $\alpha=0.1\pm0.6$. 

Further VLA observations were performed on UT 2012 August
3, using C, Ku (14\,GHz), and K (20\,GHz) bands and on UT 2012 August
5 using C band. We do not detect the supernova ($> 3\sigma$) in both
the second and third epochs in all bands.

On UT 2012 August 03, we also observed PTF\,12gzk with CARMA at a
frequency of 95\,GHz. The
observation was performed in E array and 3C446 was used as the phase
calibrator. The data were reduced using the MIRIAD software. The SN was
not detected with a $3\sigma$ upper limit of 3.6\,mJy.
The log of the cm- and mm-wavelength observations can be found in Table 1. 
\begin{table*}[!ht]

\caption{Summary of radio observations of PTF\,12gzk}\smallskip
\begin{center}
\begin{tabular}{lcll}
\hline
\noalign{\smallskip}
Day & Frequency   & Flux  & Observatory  \\
               &  [GHz] & [$\mu$Jy] &  \\
\noalign{\smallskip}
\hline
\noalign{\smallskip}

1.42 &	 4.8 &  $79\pm 16$ & VLA \\
1.42	&    7.4  &  $82\pm 15$ & VLA \\
3.33 &	 6.1& 	 $\leq 33$ & VLA \\
3.33 &	 14 &	 $\leq 42$  & VLA \\ 
3.33 &	 20	 & $\leq 90$ & VLA \\
3.34 &       95     & $\leq 3600$ & CARMA \\
5.55 &	 6.1	 & $\leq 33$  & VLA    \\

\noalign{\smallskip}
\hline
\smallskip
\end{tabular} 
\label{tab:RadioLog}
\end{center}
{\small
Notes - Day is given in UT days in 2012
August. The error in the flux is the rms error
measured in each image. 
Upper limits are $3\sigma$ limits.}            
\end{table*}

\subsection{X-ray Observation} PTF\,12gzk was observed by the X-Ray Telescope (XRT; Burrows et
 al. 2005) and the ultraviolet imaging
 telescope (UVOT; Roming et al. 2005) on the
 {\it Swift} satellite. 
XRT measurements, beginning at 13:39 UT on July 31, detected no source
at the location of PTF\,12gzk. We estimate a dead-time-corrected 
limit on the XRT count rate of $<2\times 10^{-3}$ ct s$^{-1}$. Assuming a
power-law spectrum with a photon index of 2, this corresponds to a
limit on the X-ray flux of $<7\times 10^{-14}$ erg cm$^{-2}$ s$^{-1}$ (
$L_{X}<2.8\times 10^{40}$ erg s$^{-1}$).

\subsection{A Search for High-energy Emission}

We have searched the available public archives for GRBs contemporaneous with PTF\,12gzk.  Using the inferred constraints
on the explosion date from early optical observations (see \S 2.1), we find no reported GRBs from
either the InterPlanetary Network (IPN; Hurley et al. 2010) or the 
\textit{Fermi} $\gamma$-ray Burst Monitor (GBM; Meegan et al. 2009).
Unlike the GBM, the IPN provides essentially continuous all-sky
coverage, to a limiting fluence (10\,keV -- 5\,MeV) of $S_{\gamma}
\lesssim 2 \times 10^{-6}$\,erg\,cm$^{-2}$, so we shall
adopt this value as a limit on any high-energy emission associated
with PTF\,12gzk. 

The lack of gamma-ray emission from PTF\,12gzk therefore suggests that
an associated GRB (if one exists) had to be an off-axis event.

\section{Shockwave Properties and Energetics}

\begin{figure}[!ht]
\centering
\includegraphics[width=0.45\textwidth]{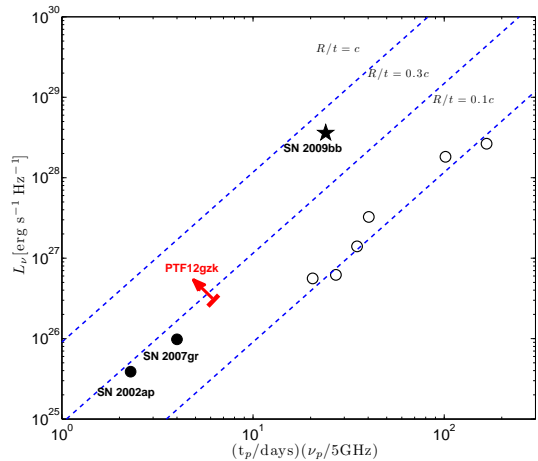}
\caption{Chevalier luminosity-peak time diagram. The open black circles are
  measurements of ``normal'' Ic SNe (Chevalier \& Fransson 2006 and
  references therein). The 5-GHz luminosity of PTF\,12gzk is $3\times 10^{26}\,
{\rm erg}$ s$^{-1}$ Hz$^{-1}$ on
UT 2012 August 1. Likely this corresponds to a post-explosion epoch of
eight days. 
The radio emission is already decaying. Therefore, the peak time is 
$\leq 7$ days. From this we conclude that PTF\,12gzk is
a high speed Ic supernova with mean ejecta velocity (of the leading edge) in excess
of 80,000 km s$^{-1}$.}
\label{fig:chevalier}
\end{figure}

The sparse data prevents us from performing any detailed modeling of
PTF\,12gzk. However, we can approximate the lower limit on the
shockwave radius, and therefore its velocity, using the formulation of
Chevalier (1998; Equation 13). Assuming equipartition, (i.e., the fraction of shockwave
energy converted to electron acceleration is equal to the fraction of
energy converted to a magnetic field), and adopting the lower limit on the
peak flux at 5\,GHz $F_{p}\geq 77$\,$\mu$Jy, yields a lower limit on the
shockwave radius of $R_s\geq 4.8\times 10^{15}$\,cm. The shockwave radius can be translated
into a lower limit on the shockwave velocity, assuming
that the time of the peak is $t\leq7$ days\footnote{The limit on the
  time of the peak is estimated based on the explosion time from
  Ben-ami et al. (2012b) and the assumption that the emission is already
  fading on UT 2012 August 1. If the emission was still rising on UT 2012
  Aug 1, we should have detected the SN on our second
  epoch of observation less than two days later}. Thus, $v_s=Rs/t\geq
80,000$\,km s$^{-1}$. This can also be graphically inferred from the 
Chevalier luminosity-peak time diagram (Figure~\ref{fig:chevalier}).

In the same
manner, we can approximate the mass-loss rate from the progenitor
prior to explosion, using equation 23 from Chevalier \& Fransson (2006)
and assuming that the CSM around the SN was deposited
by a stellar wind with a constant velocity, $v_{w}$, and a constant
mass-loss rate, $\dot {\rm M}$. Adopting the value for the fraction of energy of the
shockwave that is converted into a magnetic field $\epsilon_{\rm
  B}=0.1$, we find that the mass-loading parameter defined as
$A=\frac{\dot {\rm M}}{4\pi v_{w}}$ is
  $A\leq 6\times 10^{10} {\rm g~cm}^{-1}$. This is also consistent
 with the upper limit on the X-ray flux, assuming that any X-ray emission
 originates from inverse-Compton (see Chevlaier \& Fransson 2006;
 Soderberg et al. 2012; Horesh et al. 2012).

We note
that in our analysis so far we have assumed equipartition
($\epsilon_{e}=\epsilon_{\rm B}$). However, as previous studies
show, equipartition is not necessarily the case (Soderberg et
al. 2012; Horesh et al. 2012). Adopting a more extreme value of 
$\epsilon_e/\epsilon_{\rm B} = 500$ (motivated by the result of
Horesh et al. 2012) lowers the lower limit on the
shockwave velocity to $v_s\geq 60,000$\,km s$^{-1}$ ($v_s \propto (\epsilon_e/\epsilon_{\rm B})^{-1/19}$) .

At this point we also would like to note that the change in radio flux
between the first and second epochs is significantly larger
than would be predicted by a $t^{-1}$ decay (for a shockwave with
a constant velocity and electron energy power-law index of $p=3$). However, the flux that we measure may be
effected by interstellar scattering and scintillation (ISS) and therefore
may be intrinsically lower.

Taking into account the ISS modulation index, the intrinsic flux of
PTF\,12gzk can be as low as $45$\,$\mu$Jy. In this case the shockwave
velocity lower limit reduces from $80,000$ km s$^{-1}$ to $60,000$ km s$^{-1}$. Moreover, the
difference between the observed reduced flux on the first epoch and
the flux limit on the second epoch, is consistant (within the
errors) with the expected
change in the Chevalier (1998) synchrotron emission model.

We find that the internal minimum energy of the 
emitting material is $E_{\rm min}=7.3\times 10^{45}$
erg (See equation $7$ in Horesh et al. 2012). The energy of an ejecta velocity component in a pure explosion is given by 
\begin{equation}
E(v)=E_0\left(\frac{v}{v_0}\right)^{-5}
\end{equation}
where $v_0$ is the maximum velocity of the ejecta and
$E_0$ is the energy carried by it. Considering a radiative progenitor star (polytropic index of $n=3$), i.e., a Wolf-Rayet (WR) or a
blue supergiant (BSG) star, and substituting the equations of Matzner
\& Mckee (1999)\footnote{Equations 32 \& 37 in
  Matzner \& Mckee (1999)} for $v_0$ and $E_0$ (see also Nakar \&
Sari 2010 for the normalizations that are used here)
into the above equation, the energy\footnote{This is the energy
  of the fast-moving ejecta originating from the outer layer of
  the progenitor star, in contrast to the energy carried by the bulk ejecta 
  that originates from the inner parts of the star.} can be now expressed as
\begin{eqnarray}
E_{ej}(v)&=&2.3\times 10^{52} \left(\frac{M}{
M_{\odot}}\right)^{-2.46}\left(\frac{E}{10^{51}
{\rm erg}}\right)^{3.48}\nonumber\\
& &\times \left(\frac{v}{10,000~{\rm km~s}^{-1}}\right)^{-5}  
\end{eqnarray}

Plugging our velocity lower limit, $v_{0}\geq 80,000$ km s$^{-1}$,
into the above equation and adopting the ejecta mass and explosion energy
values from Ben-Ami et al. (2012b) ($M_{ej}=7.5\, M_{\odot}$ and
$E=7\times 10^{51}$ erg) yields $E_{ej}\approx 4.3\times 10^{48}$
erg. Therefore, there is more than enough energy in the ejecta to
account for the shockwave minimum energy that we measure. 

\section{Comparison with a Sample of Radio-detected SNe Ib/c}

The two most comparable SNe Ic that also exhibit high shockwave
velocities, but are not relativistic, 
are SN\,2002ap (Berger et
al. 2002) and SN\,2007gr (Soderberg et al. 2010). Both of these sources are believed to be ordinary core-collapse explosions (cf., Paragi et al. 2010). Both SNe had radio
emission that evolved relatively fast, with SN\,2002ap peaking at
$t\sim 2-3$ days (4.86\,GHz) and SN\,2007gr peaking at $t\sim 3-4$\,days
(8.46\,GHz), after explosion. The radio analysis by both Berger et al. (2002) and Soderberg et
al. (2010) suggest shockwave velocities of $\approx 0.2-0.3c$. For
PTF\,12gzk, as seen in Figure~\ref{fig:chevalier}, the lower limit is about $0.3c$,
similar to the velocities of SN\,2002ap and SN\,2007gr. However, the
two latter SNe had fainter radio luminosities by factors of $\sim 10$ and $\sim 3$ compared with PTF\,12gzk,
respectively. The brighter radio emission from PTF\,12gzk may suggest
that the shockwave velocity is even higher in this case.

The optical spectra of SN\,2002ap, SN\,2007gr and PTF\,12gzk show significant
difference. The spectra of SN\,2002ap (Type Ic-BL)
exhibit broad lines formed by emission from ejecta with a broad range of
velocities. The velocity of the line centers suggests a bulk velocity of $\sim
38,000$\,km s$^{-1}$ 7 days prior to optical peak (B-band) that decreased to $15,000$\,km s$^{-1}$ at
peak (Gal-Yam, Ofek \& Shemmer 2002). In the case of SN\,2007gr (normal Ic), there
are no broad lines and the
absorption line blueshifts suggest a velocity of less than 13,000 km s$^{-1}$
(Hunter et al. 2009). PTF\,12gzk, in contrast, does not show
broad lines\footnote{PTF\,12gzk has narrow lines with a dispersion
  lower by a factor of $\sim 4$ compared to broad-lined SNe
  Ic (see Ben-Ami et al. 2012b for details).} as does SN\,2002ap, but based on
line blueshifts, it does exhibit high expansion velocities of $\sim
35,000$\,km s$^{-1}$, a factor of $\sim 3$ higher than the velocity of SN\,2007gr. 
This higher velocity is consistent with our inference that the shock
velocity traced by the radio observations is larger than $0.2-0.3c$.

Another property that PTF\,12gzk has in common with SN\,2002ap and
SN\,2007gr is the inferred mass-loading parameter of the SN progenitor. In
Table~\ref{tab:Astar} we list the normalized mass-loading parameter, $A_{\star}$, values of these three
SNe and a few other Ib/c SNe, where $A_{\star}=A/5\times
10^{11}{\rm g~cm}^{-1}$. PTF\,12gzk, SN\,2002ap and SN\,2007gr
all have $A_{\star}$ values which are at least a factor of $10$ lower
than the values for the rest of the listed SNe. This is an
interesting clue worthy of further investigation, that may suggest more compact
progenitors for these objects, driving faster winds.

\begin{table*}[!ht]

\caption{A$_{\star}$ values of PTF\,12gzk in comparison with other SNe Ib/c}
\smallskip
\begin{center}
\begin{tabular}{lrll}
\hline
\noalign{\smallskip}
SN           &  Type   & Velocity (optical) & A$_{\star}$   \\
               &            & km\,s$^{-1}$      &                   \\
\hline
1983N      & Ib        & 18,200              &     1.15              \\
1990B      & Ic        & 17,700             &      2.0             \\
1994I       & Ic         & 17,500              &      2.8             \\
2003L      & Ic         & 12,000              &       34            \\
2002ap    & Ic-BL    & 38,000                &     0.04              \\
2007gr     & Ic         & 13,000               &      0.06            \\
PTF\,12gzk & Ic       & 35,000              &      $\leq  0.12$            \\
\noalign{\smallskip}

\noalign{\smallskip}
\hline
\smallskip
\end{tabular} 
\label{tab:Astar}
\end{center}
{\small
Notes - The optical velocities are taken from  Wheeler \& Harkness (1990), Matheson et al. (2001), Millard et al. (1999),
Matheson et al. (2003),
Gal-Yam, Ofek, \& Shemmer (2002), Hunter et al. (2009) and Ben-Ami et
al. (2012), respectively. The $A_{\star}$ values are from
Chevalier \& Fransson (2006) when equipartition is assumed and we
adopt the value of $\epsilon_{\rm B}=0.1$.}     
\end{table*}

\section{Summary and Future Implications}

We observed PTF\,12gzk in mm-, cm-wavelength and X-rays, starting eight days
after explosion. Our radio observations reveal a rapidly declining
emission that can be explained by a very fast $\geq 80,000$ km s$^{-1}$
shockwave ploughing through an optically thin CSM, that was deposited by  a
stellar wind, prior to the explosion. We also calculate the limit on
the mass-loading parameter $A\leq 6\times 10^{10}\, {\rm g~cm}^{-1}$, which is 
consistent with our X-ray observations. Observations of PTF\,12gzk can be explained
by an ordinary pure explosion supernova, and do not require an engine-driven
event. However, its inferred high-velocity shockwave (based on 
our radio analysis) together with the high velocity of the bulk of the
ejecta (based on the optical spectra; Ben-Ami et al. 2012), may indicate that
it is an intermediate event between a ``normal'' SN Ic and a GRB-SN like
event.

The fraction of rapidly declining (peak emission at 5\,GHz in less than 10
days) radio SNe Ic is not well constrained. While the discovery rate of SNe Ic by 
PTF is $\sim 25$ per year on
average, many of these past events were discovered many days after explosion. 
Recently, PTF has adopted a higher survey
cadence that enables the discovery of SNe within one day of explosion. 
Figure 1 in Soderberg et al. (2010) reveals that most of the radio observations of SNe Ic took
place at late times. Therefore, most rapidly declining events like PTF\,12gzk would have been missed by
current observing strategies. 
Thus, it is possible that the average shockwave velocities of
$0.1-0.15c$ observed in normal SNe Ic may not represent the
average velocity of the general SN Ic population.

The more improved sensitivity of the VLA offers the opportunity to detect radio emission from high-velocity SNe Ic out to larger
distances. Past studies probed only nearby ($\leq 10$ Mpc) rapidly evolving SNe Ic such as
SN\,2007gr and SN\,2002ap, due to limited sensitivity. Given an rms
of $10$\,$\mu$Jy in C band,  the detection horizon for SNe Ic with a
shockwave velocity of $0.3c$ and with an $A_{\star}$ value similar to
that of PTF\,12gzk, is $70$\,Mpc, and $20$\,Mpc even for SNe with an
$A_{\star}$ value lower by a
factor of $10$ than the one of PTF\,12gzk. Relativistic events, such
as SN\,2009bb, will be detectable out to a distance of $100 - 300$
Mpc, if they will have an $A_{\star}$ value similar to that of PTF\,12gzk or
even lower.

Another possible strategy to better study rapidly evolving radio SN is
to conduct the observations at lower frequencies (e.g., 1.4\,GHz). 
The radio emission at 1.4\,GHz for events such as PTF\,12gzk,
SN\,2007gr, and SN\, 2002ap is expected to peak after more than $10$\,days
after explosion. 
Therefore, future surveys at low frequencies such as Apertif
(Verheijen et al. 2008), and ASKAP (Johnston et al. 2008) have
the potential to discover and follow many such SNe.

\begin{figure}[htbp] 
   \centering
   \includegraphics[width=3.3in]{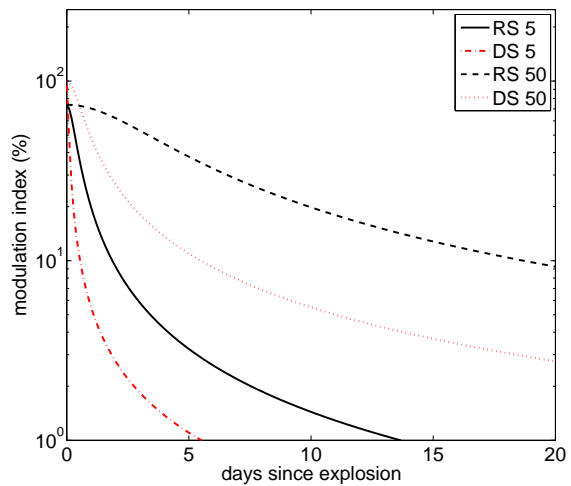}
   \caption{\small
   Modulation index for observations in the 5\,GHz band of a 
   supernova at a distance of 5\,Mpc and 50\,Mpc and expanding with
   a blast wave speed of $10^{5} {\rm km~s}^{-1}$. We assume log(SM)=$-3.63$. The scattering strength is above unity and so both refractive  broad-band
   scintillation (RS) and diffractive (narrow band) scintillations
   (DS) are seen. ISS calculations are based on Walker (1998).
   }
\label{fig:5GHzISS}
\end{figure}

One of the main challenges in studying radio emitting SNe Ic at larger
distances by either the VLA or future low-frequency surveys is ISS. 
As we probe further out, the angular size of the source decreases and
therefore the effect of ISS increases. Nature, at this point, is
becoming the main source of noise due to ISS. According to Figure~\ref{fig:5GHzISS}, a
SN-CSM shockwave, expanding at a velocity of $0.3c$, for example, will
still exhibit more than $20\%$ variability due to ISS at a distance
of 50\,Mpc, after 10 days. At the same time, one can take the approach
of undertaking high-frequency (mm-wavelength) observations which are less susceptible to
ISS modulation. Since high-frequency emission in PTF\,12gzk-like
events is expected to peak at $\leq 1$\,day after explosion, this
approach will be efective only if immediate observations are initiated
after the detection of a SN. 

To summarize, it is now clear that there exists a phase space of fast SNe Ic
that is still relatively unexplored. The improved capabilities of radio observatories combined with early
discoveries by optical transient surveys will allow us to
probe this new population of radio SNe. Such studies will provide a
better understanding of the properties of the progenitor stars and
their explosion energies, complementing previous work based on late-time observations.

\section*{Acknowledgments}

We thank the VLA, {\it Swift}, and CARMA staff for promptly scheduling this target
of opportunity. The National Radio Astronomy Observatory is a facility
of the National Science Foundation operated under cooperative
agreement by Associated Universities, Inc. This work made use of data supplied by the UK Swift
Science Data Centre at the University of Leicester. PTF is a fully-automated, wide-field survey aimed at a systematic exploration of explosions and variable phenomena in optical wavelengths. The participating institutions are
Caltech, Columbia University, Weizmann Institute of Science, Lawrence
Berkeley Laboratory, Oxford and University of California at
Berkeley. The program is centered on a 12Kx8K, 7.8 square degree CCD
array (CFH12K) re-engineered for the 1.2-m Oschin Telescope at the
Palomar Observatory by Caltech Optical Observatories. Photometric
follow-up is undertaken by the automated Palomar 1.5-m
telescope. The research of A.G. is supported by grants from the ISF, BSF, GIF and Minerva,
the EU/FP7 via an ERC grant, and the Kimmel award for innovative investigation.

{}


\begin{thebibliography}{}

\bibitem[\protect\citeauthoryear{Ben-Ami et 
al.}{2012}]{2012ApJ...760L..33B} Ben-Ami S., et al., 2012, ApJ, 760,
L33 

\bibitem[\protect\citeauthoryear{Ben-Ami et 
al.}{2012}]{2012ATel.4297....1B} Ben-Ami S., et al., 2012, ATel, 4297,
1 
\bibitem[\protect\citeauthoryear{Berger, Kulkarni, 
\& Chevalier}{2002}]{2002ApJ...577L...5B} Berger E., Kulkarni S.~R.,
Chevalier R.~A., 2002, ApJ, 577, L5 

\bibitem[\protect\citeauthoryear{Burrows et 
al.}{2005}]{2005SSRv..120..165B} Burrows D.~N., et al., 2005, SSRv, 120, 
165 

\bibitem[Chevalier 
\& Soderberg(2010)]{2010ApJ...711L..40C} Chevalier, R.~A., \&
Soderberg, A.~M.\ 2010, \apjl, 711, L40 

\bibitem[\protect\citeauthoryear{Chevalier 
\& Fransson}{2006}]{2006ApJ...651..381C} Chevalier R.~A., Fransson C.,
2006, ApJ, 651, 381 

\bibitem[Chevalier(1998)]{1998ApJ...499..810C} Chevalier, R.~A.\ 1998, 
\apj, 499, 810 

\bibitem[\protect\citeauthoryear{Chevalier}{1982}]{1982ApJ...259..302C} 
Chevalier R.~A., 1982, ApJ, 259, 302 

\bibitem[\protect\citeauthoryear{Cordes 
\& Lazio}{2001}]{2001ApJ...549..997C} Cordes J.~M., Lazio T.~J.~W.,
2001, ApJ, 549, 997 

\bibitem[\protect\citeauthoryear{Frail et al.}{2001}]{2001ApJ...562L..55F} 
Frail D.~A., et al., 2001, ApJ, 562, L55 

\bibitem[\protect\citeauthoryear{Gal-Yam, Ofek, 
\& Shemmer}{2002}]{2002MNRAS.332L..73G} Gal-Yam A., Ofek E.~O.,
Shemmer O., 2002, MNRAS, 332, L73 


\bibitem[\protect\citeauthoryear{Hunter et 
al.}{2009}]{2009A&A...508..371H} Hunter D.~J., et al., 2009, A\&A, 508, 371 


\bibitem[\protect\citeauthoryear{Hurley et al.}{2010}]{2010AIPC.1279..330H} 
Hurley K., et al., 2010, AIPC, 1279, 330 

\bibitem[\protect\citeauthoryear{Johnston et 
al.}{2008}]{2008ExA....22..151J} Johnston S., et al., 2008, ExA, 22, 151 


\bibitem[Law et al.(2009)]{2009PASP..121.1395L} Law, N.~M., et al.\ 2009, 
\pasp, 121, 1395 

\bibitem[\protect\citeauthoryear{Mason et al.}{2004}]{2004SPIE.5165..277M} 
Mason K.~O., Breeveld A., Hunsberger S.~D., James C., Kennedy T.~E., Roming 
P.~W.~A., Stock J., 2004, SPIE, 5165, 277 

\bibitem[\protect\citeauthoryear{Matheson et 
al.}{2001}]{2001AJ....121.1648M} Matheson T., Filippenko A.~V., Li W., 
Leonard D.~C., Shields J.~C., 2001, AJ, 121, 1648 

\bibitem[\protect\citeauthoryear{Matheson et 
al.}{2003}]{2003GCN..1846....1M} Matheson T., Challis P., Kirshner R.~P., 
Garnavich P.~M., 2003, GCN, 1846, 1 

\bibitem[\protect\citeauthoryear{Matzner 
\& McKee}{1999}]{1999ApJ...510..379M} Matzner C.~D., McKee C.~F.,
1999, ApJ, 510, 379 

\bibitem[\protect\citeauthoryear{Meegan et al.}{2009}]{2009ApJ...702..791M} 
Meegan C., et al., 2009, ApJ, 702, 791 

\bibitem[\protect\citeauthoryear{Millard et 
al.}{1999}]{1999ApJ...527..746M} Millard J., et al., 1999, ApJ, 527,
746 

\bibitem[\protect\citeauthoryear{Nakar 
\& Sari}{2010}]{2010ApJ...725..904N} Nakar E., Sari R., 2010, ApJ,
725, 904 

\bibitem[\protect\citeauthoryear{Pian et al.}{2000}]{2000ApJ...536..778P} 
Pian E., et al., 2000, ApJ, 536, 778 


\bibitem[\protect\citeauthoryear{Soderberg et 
al.}{2010}]{2010Natur.463..513S} Soderberg A.~M., et al., 2010, Natur, 463, 
513 

\bibitem[\protect\citeauthoryear{Soderberg et 
al.}{2010}]{2010ApJ...725..922S} Soderberg A.~M., Brunthaler A., Nakar E., 
Chevalier R.~A., Bietenholz M.~F., 2010, ApJ, 725, 922 

\bibitem[\protect\citeauthoryear{Soderberg et 
al.}{2012}]{2012ApJ...752...78S} Soderberg A.~M., et al., 2012, ApJ, 752, 
78 

\bibitem[Rau et al.(2009)]{2009PASP..121.1334R} Rau, A., et al.\ 2009, 
\pasp, 121, 1334 


\bibitem[\protect\citeauthoryear{Roming et al.}{2005}]{2005SSRv..120...95R} 
Roming P.~W.~A., et al., 2005, SSRv, 120, 95 

\bibitem[\protect\citeauthoryear{Verheijen et 
al.}{2008}]{2008AIPC.1035..265V} Verheijen M.~A.~W., Oosterloo T.~A., van 
Cappellen W.~A., Bakker L., Ivashina M.~V., van der Hulst J.~M., 2008, 
AIPC, 1035, 265 

\bibitem[\protect\citeauthoryear{Walker}{1998}]{1998MNRAS.294..307W} Walker 
M.~A., 1998, MNRAS, 294, 307 

\bibitem[\protect\citeauthoryear{Walker}{2001}]{2001MNRAS.321..176W} Walker 
M.~A., 2001, MNRAS, 321, 176 

\bibitem[\protect\citeauthoryear{Weiler et 
al.}{2002}]{2002ARA&A..40..387W} Weiler K.~W., Panagia N., Montes
M.~J., Sramek R.~A., 2002, ARA\&A, 40, 387 

\bibitem[\protect\citeauthoryear{Wheeler 
\& Harkness}{1990}]{1990RPPh...53.1467W} Wheeler J.~C., Harkness
R.~P., 1990, RPPh, 53, 1467 

\end{thebibliography}
\end{document}